\documentstyle[epsf]{mn}

\def\figdir{.}

\def\figname#1{\figdir/#1}

\title{CO emission from high redshift}
\author[Gnedin, Silk, \& Spaans]
{Nickolay Y.\ Gnedin$^1$, Joseph Silk$^2$,
  and Marco Spaans$^3$\\
  $^1$Center for Astrophysics and Space Astronomy, 
University of Colorado, Boulder, CO 80309, USA; gnedin@casa.colorado.edu\\
  $^2$Astrophysics, Department of Physics, Oxford
                  University, Keble Road, Oxford, OX1 3NP, UK\\
  $^3$Kapteyn Institute, Groningen, 9700 AV, The Netherlands}
\date{}
\pagerange{\pageref{firstpage}--\pageref{lastpage}}
\pubyear{}

\begin{document}

\def\dim#1{\mbox{\,#1}}

\label{firstpage}

\maketitle

\begin{abstract}
Future observations with ALMA will be able to detect star-forming primordial galaxies, and perhaps 
even their dwarf spheroidal satellites, in CO
emission lines at redshifts approaching 10. These observations will
compliment other tools designed to study the dawn of galaxy formation,
such as NGST and FIRST.
\end{abstract}

\begin{keywords}
 cosmology: theory -- galaxies: evolution -- galaxies: starburst -- 
ISM: molecules
\end{keywords}

\section{Introduction}

Modern cosmology continues to push the boundaries of the known universe to
higher and higher redshifts. We are designing telescopes capable of detecting the very first objects that formed in
the universe.
The real breakthrough will
occur in the next decade: the New Generation Space Telescope (NGST), the
The Far IR and Submillimiter Space Telescope (FIRST), and 
the Atacama Large Millimeter Array (ALMA) will allow us to reach cosmological
redshifts in excess of 10.
ALMA is especially important, because it can observe CO emission
from primordial galaxies redshifted into the millimeter band. Because
the Cosmic Microwave Background (CMB) temperature increases toward the past,
higher rotational levels of the CO molecule are populated at high redshifts
\cite{ss,s99,bea00}, 
resulting in a  large negative K-correction. The effect
is so large that a star-forming galaxy will appear equally bright at $z=5$ 
and at $z=10$. 

At lower redshift, CO emission from galaxies and quasars has already been
detected 
\cite{sea92,bea94,oea96,gea97,sea97,bea98,fea98,dea99,fea99,aea00,pea00,pea01}.
These results provide information on the kinematics and energetics of the
star-forming (molecular) ISM and thereby complement observations of the stellar
light component in primordial systems.

In this paper we show
that ALMA, with its significantly higher sensitivity, will be able to
image primordial galaxies  to redshift 10 and even beyond.
It is conceivable that it will even be able to detect dwarf
spheroidal satellites of primordial galaxies and their tidal tails
(which should be in abundance at high redshift, where the merger rate
is much higher than in the local universe).

\section{Method}

\subsection{Simulations}

We use the cosmological simulations of reionization
reported in Gnedin \shortcite{me}. The simulations include 3D radiative
transfer (in an approximate implementation) and other physical ingredients
required for modeling the process of cosmological reionization. 

\begin{table}
\caption{Simulation Parameters}
\label{sim}
\begin{tabular}{@{}ccccc}
Run & 
$N$ & 
Box size & 
Mass res. & 
Spatial res. \\
A & $128^3$ & $4h^{-1}{\rm\,Mpc}$ & 
$10^{6.6}\dim{M}_{\sun}$ &$1.0h^{-1}{\rm\,kpc}$ \\
B & $64^3$ & $2h^{-1}{\rm\,Mpc}$ & 
$10^{6.6}\dim{M}_{\sun}$ &$1.5h^{-1}{\rm\,kpc}$ \\
\end{tabular}
\end{table}
Two simulations of a representative CDM+$\Lambda$ cosmological 
model\footnote{With the following cosmological parameters: $\Omega_0=0.3$,
$\Omega_\Lambda=0.7$, 
$h=0.7$, $\Omega_b=0.04$, $n=1$, $\sigma_8=0.91$, where the amplitude and
the slope of the primordial spectrum are fixed by the COBE and cluster-scale
normalizations.}
were performed with the parameters specified in Table \ref{sim}.
Both simulations were stopped at $z=4$ because at this time the rms density 
fluctuation in the computational box is about 0.25, and at later times the 
box ceases to be a representative region of the universe.

The two simulations from Table \ref{sim} allow us to investigate the 
sensitivity of our results to the inevitably missing small-scale and large-scale power.
The difference between the two runs can be interpreted as the theoretical
uncertainty in our calculations, given a cosmological model. Clearly,
our results will be different for different assumptions about cosmological 
parameters.

\subsection{Calculating CO emission from early galaxies}

The code described in Spaans \shortcite{s96} and applied as in 
Silk \& Spaans \shortcite{ss}
has been used to rerun the models presented in \cite{ss} with
the latest atomic and molecular collision and chemistry data. These models
use the Orion molecular
cloud and its so-called bar region as being representative of a region of
active star formation. The star formation rate in $M_\odot$ yr$^{-1}$ of
a fiducial  model galaxy is then related to the total number of Orion-like star
formation sites through division by the average star formation rate of
the Orion region, $\sim 3\times 10^{-4}$ $M_\odot$ yr$^{-1}$ \cite{h97}.
In Silk \& Spaans \shortcite{ss} it has been shown that the CMB becomes an
important source of excitation at high redshift because of the fortuitous
coincidence between the CO level spacing and the $1+z$ increase in the CMB
temperature. This causes galaxies at $z=5$ and $z=10$ to be observable at
similar flux density levels, provided they in fact are present.
It has been assumed that the Orion-like regions responsible for the star
formation activity occur throughout the model
galaxy, and are not all confined to the inner few 100 pc as in 
Combes, Maoli, \& Omont \shortcite{cmo}
This assumption decreases the mean optical depth of the CO lines and is most
likely to hold at high ($z>3$) redshifts,
when galaxies are still being built up through mergers and accretion.

In order to compute the spectrum of CO emission as a function of
wavelength, for a given bandwidth $\lambda_1<\lambda<\lambda_2$ and
a given transition $J\rightarrow J-1$, we identify a range of cosmological
scale factors
$a_1<a<a_2$ that correspond to our bandwidth. This range of scale factors 
in turn corresponds to the range of comoving distances $x_1<x<x_2$.
However, due to periodic
boundary conditions adopted in the simulations, we cannot always model this range
of comoving distances directly - if it is large enough, it will correspond to
more than one box size. If we simply stack a sufficient number of simulation
boxes together, we will get an unphysical result due to periodicity. In 
order to break this periodicity, we use the approach described in
Gnedin \& Jaffe \shortcite{gj}: we randomize the neighboring boxes by
randomly flipping, transposing, and shifting each of the periodic images of
the computational box.

\section{Results}

\begin{figure}
\epsfxsize=1.0\columnwidth\epsfbox{\figname{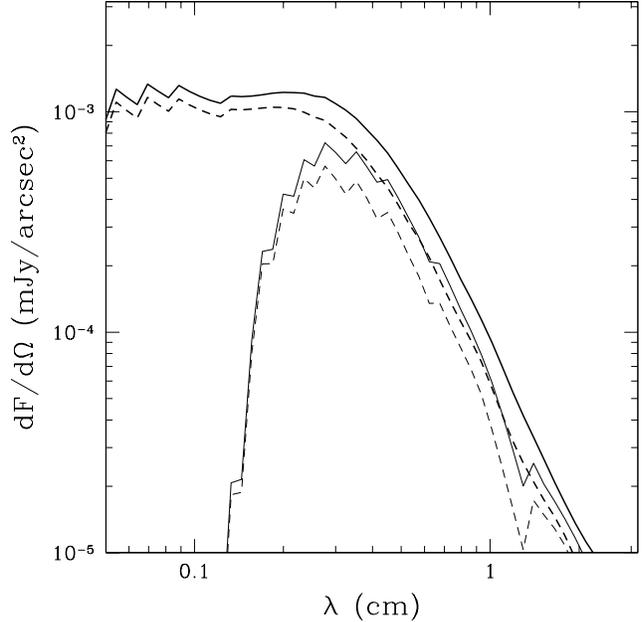}}
\caption{The mean CO flux density
as a function of wavelength from two simulations:
the small simulation box (dashed line) and the large box (solid line),
as described in Table 1. Thin lines
show the calculation with the star formation rate set to zero after the
final redshift of simulation ($z=4$); bold lines are for the case when
the star formation rate is assumed constant for $z<4$. The difference between
the two sets of lines is our theoretical uncertainty due to the final
size of the simulation box.}
\label{figAE}
\end{figure}
Figure \ref{figAE} serves to illustrate the uncertainty of our calculations 
due to the finite size of the computational box and finite numerical 
resolution, as measured by the difference between the two simulations A and B.
In addition, since both simulations were stopped at $z=4$, a contribution from
later redshifts cannot be included. In order to estimate the effect of this
contribution, we calculated the CO emission for two cases: no star formation 
after $z=4$, and constant star formation after $z=4$. The difference between
those two cases quantifies the uncertainty due to the finite value for the
final redshift of our simulation.

For $\lambda<0.3\dim{cm}$ our calculation is not reliable even in a 
qualitative sense (to within a factor of 2). At higher wavelengths finite
numerical resolution still prevents us from achieving better than about 50\% 
accuracy. More than that, since the star formation rate in our simulations
is normalized to the observed value at $z=4$, which is in turn uncertain
to at least a factor of two \cite{nce,sea}, our results in general are
uncertain to a factor of two to three. However this is quite sufficient for our 
purpose, which is to emphasize the possibilities rather than to
make some
definite predictions. Because ALMA
will not be commissioned until about 2010, theorists have plenty of time to
improve upon our calculations and come up with more definitive predictions.

\begin{figure}
\epsfxsize=1.0\columnwidth\epsfbox{\figname{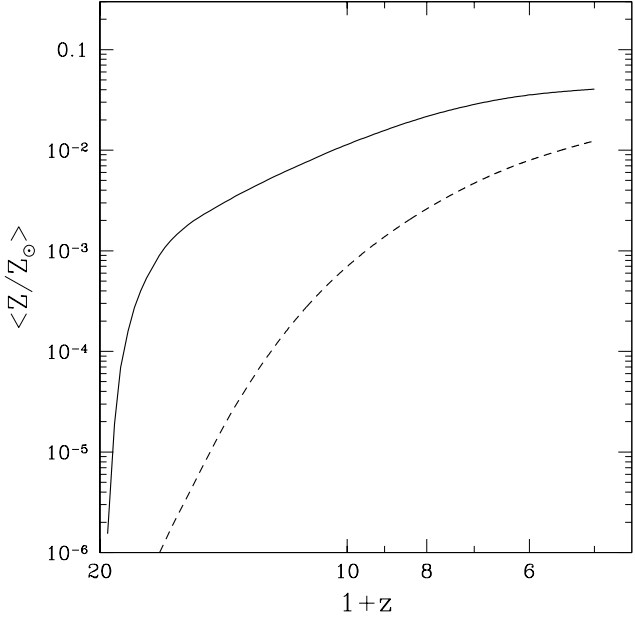}}
\caption{Evolution of the mean mass-weighted metallicity of
the gas (dashed line) and stars (solid line).
}
\label{figZA}
\end{figure}
Figure \ref{figZA} shows the evolution of the mean mass-weighted
metallicity of the
gas and stars in our large simulation (run A). One can see that stars quickly
gain a metallicity of 3\% solar by $z\sim15$, and then gain another order of
magnitude on average by the end of the simulation at $z=4$. The metallicity of
the gas is always lower than stellar, but increases more rapidly. The 
decrease in stellar metallicities 
at higher redshifts slightly mitigates the increase in the CO emission due to
higher CMB temperature, however it is not sufficient to completely remove
the negative K-correction, and thus high redshift star forming galaxies
should be considerably brighter than their low redshift counterparts.

\begin{figure}
\epsfxsize=1.0\columnwidth\epsfbox{\figname{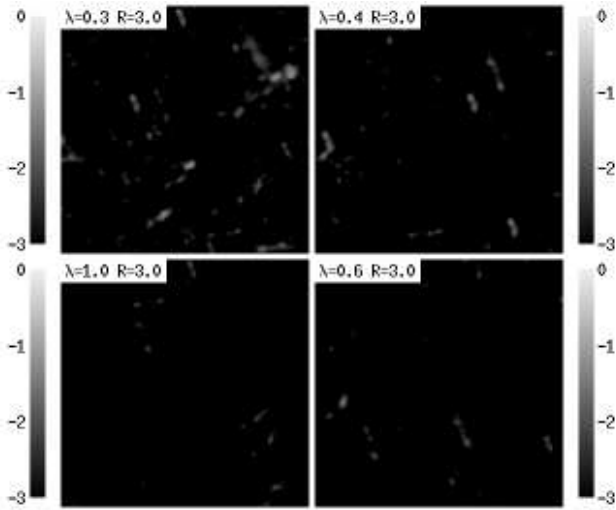}}
\caption{A $2\times2$ degree image of the sky at four different wavelength
($\lambda=0.3$, $0.4$, $0.6$, and $1\dim{cm}$ clockwise from the top left
panel, as labeled in the figure) taken with a frequency resolution 
$R\equiv \lg(\lambda/\Delta\lambda)=3$. Side bars give the correspondence 
between the color and decimal log of flux density in mJy per square arcsecond.
}
\label{figIL}
\end{figure}
In Figure \ref{figIL} we show our main result: the 
4 square degrees of the sky (an image of our
computational box) at four different wavelengths.
The peak of CO emission corresponds to a broad range of transitions 
with values in the range of $J=5$ to 12 \cite{ss}. 
This means that several epochs
correspond to a single observed wavelength: $\lambda=0.3\dim{cm}$ maps
redshifts from about 5 to 13, $\lambda=0.6\dim{cm}$ corresponds to redshifts
from about 10 to 30, and $\lambda=1.0\dim{cm}$ includes everything
from $z=20$ to $z=50$. This is somewhat unfortunate, because it means that
a simple image at a given wavelength can only provide information about
star formation over a range of redshifts. On the other hand, primordial
star forming galaxies have relatively narrow velocity ranges, which means that
by observing the CO emission in a narrow wavelength band, it is still possible
to disentangle different epochs. 

\begin{figure}
\epsfxsize=1.0\columnwidth\epsfbox{\figname{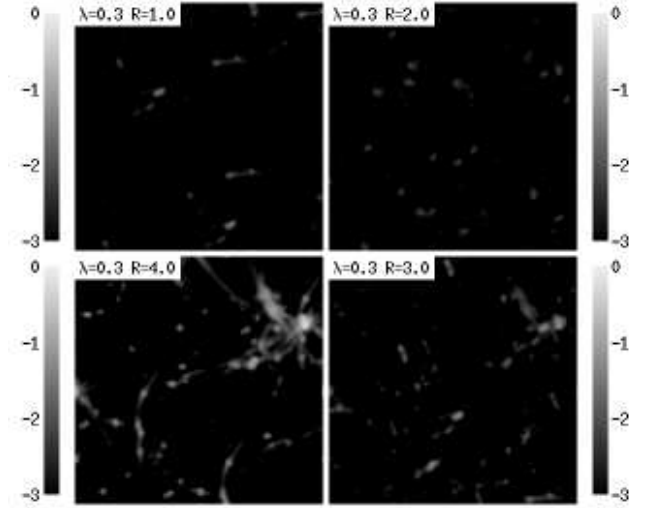}}
\caption{A $2\times2$ degree image of the sky at 
$\lambda=0.3\dim{cm}$ taken with four different wavelength resolutions
$R\equiv \lg(\lambda/\Delta\lambda)=1$, $2$, $3$, and $4$,
clockwise from the top-left panel (as labeled in the figure).
Side bars give the correspondence 
between the color and decimal log of flux density in mJy per square arcsecond.
}
\label{figIR}
\end{figure}
This is illustrated in Figure \ref{figIR}, where
we show the same wavelength but at four different spectral resolutions. With
a resolution $\lambda/\Delta\lambda=10^{4}$ we can eliminate most of the
emission from different
redshifts and obtain an image of a single primordial galaxy (lower left panel
in Fig.\ \ref{figIR}). Notice that not only the galaxy itself is measurable
by ALMA, but also its satellites - young dwarf spheroidals - are clearly
visible in the image. A fortunate circumstance -  an enhancement in population
of high $J$ levels due to higher CMB temperature - may allow ALMA even to
see dwarf spheroidals being tidally disrupted in the vicinity of larger
galaxies and to map their tidal tails (although this will be a very difficult
observation). And, of course,
the high redshift picture is much more spectacular than its local equivalents
because the merging rate at $z\sim 10$ is some 500 times higher than in
the local universe.

\begin{figure}
\epsfxsize=1.0\columnwidth\epsfbox{\figname{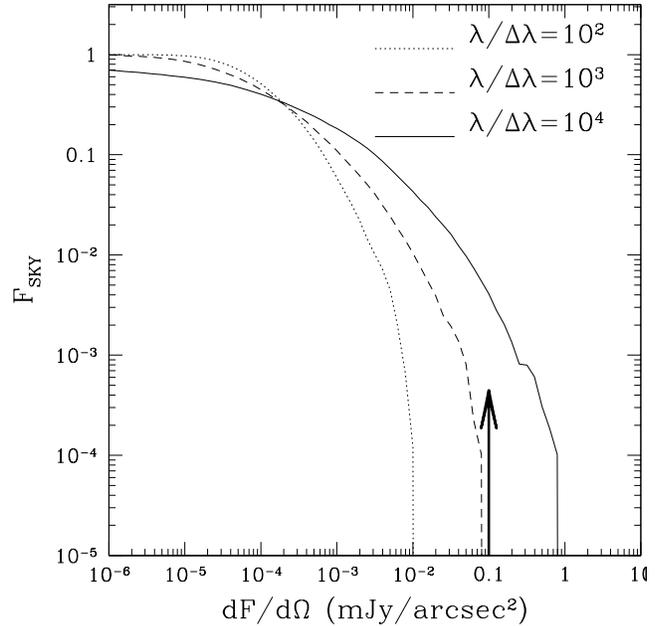}}
\caption{Fraction of the sky above a given flux density for three different
spectral resolutions at $\lambda=0.3\dim{cm}$. The black arrow shows ALMA's
sensitivity per $1\dim{arcsec}^2$ pixel for a 100 hour integration (or,
equivalently, 1 hour integration on a $10\dim{arcsec}^2$ source).
}
\label{figHF}
\end{figure}
\begin{figure}
\epsfxsize=1.0\columnwidth\epsfbox{\figname{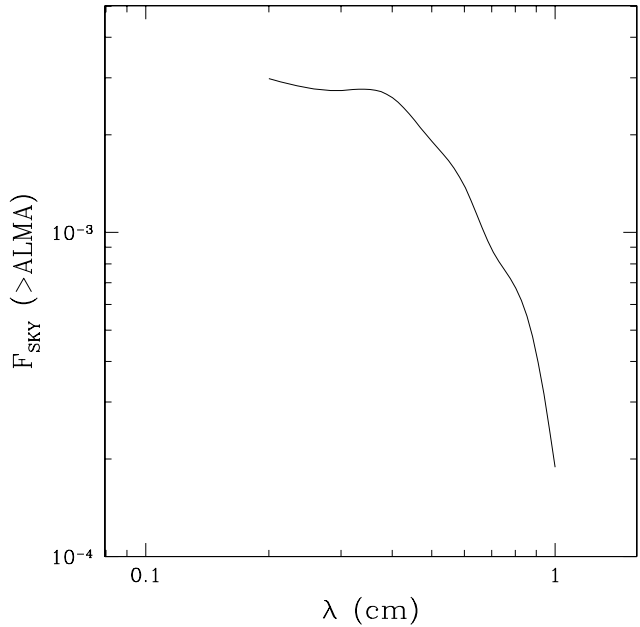}}
\caption{Fraction of the sky above the ALMA sensitivity as a function
of the wavelength. We only show wavelengths $\lambda>0.2\dim{cm}$
since at shorter wavelengths our calculations become unreliable,
as shown by Fig.\ \protect{\ref{figAE}}.
}
\label{figHL}
\end{figure}
The distribution of flux density on the sky is shown in Figure \ref{figHF}.
We also show a sensitivity limit for ALMA for a typical observation. 
While ALMA will be just short of measuring typical objects with a spectral
resolution of $10^3$, at a resolution of $10^4$ it will be able not only to
detect primordial galaxies but also to take their images. High spectral
resolution is also required to separate emission from different redshifts,
so this mode appears to be the most promising for high redshift observations
with ALMA. 

Figure \ref{figHL} summarizes our results by giving the fraction
of the sky above the ALMA sensitivity as a function of wavelength. ALMA
will have an easy time observing primeval star-forming galaxies at
$z\sim5-10$, but will not be able to see much beyond that because of the
strong decrease in stellar metallicities and star formation rates at
higher redshifts. The latter statement is, of course, strongly dependent
on the cosmological model - models with large amounts of small scale power
will have star formation commencing earlier.

\section{Conclusions}

We have demonstrated that future observations with ALMA will be able to detect
star-forming primordial galaxies, and even their dwarf spheroidal satellites,
in CO
emission lines, mostly due to a large negative K-correction. High spectral
resolution observations are required to both separate contributions
from several objects at different redshifts all emitting at the same 
wavelength from different rotational levels, and to increase the 
signal-to-noise. Direct imaging will be possible for galaxies up to
redshifts approaching 10. Unfortunately, the expected decrease in the star
formation rates and stellar metallicities at higher redshifts will make
observations of the era before $z=10$ extremely difficult.

A simple possible strategy for these observations might include a shallow
large area survey to identify early forming massive galaxies (rare peaks), 
with subsequent targeted deep observations of their environments. In this
case it is possible to observe the same region on the sky
at several wavelength bands spaced around several subsequent
CO emission lines, in order to get a linear increase in the signal-to-noise.

Broadly speaking, CO emission lines, boosted by higher CMB temperature 
in the dawn of galaxy formation, will allow millimeter observations
with ALMA and other instruments not only to  complement observations
with the Next Generation Space Telescope (NGST), but, perhaps, to directly
compete with the NGST in the race for the highest observable redshifts.

This work was partially supported by National Computational Science
Alliance under grant AST-960015N and utilized the SGI/CRAY Origin 2000 array
at the National Center for Supercomputing Applications (NCSA).

\label{lastpage}

\end{document}